\documentclass[11pt, a4paper]{article}

\usepackage{cite}
\usepackage{amsmath, amsthm,amssymb,mathtools,url ,amssymb,upgreek,slashed,amsthm,braket}
\usepackage{amsfonts}
\usepackage[normalem]{ulem}
\usepackage[utf8]{inputenc}
\usepackage[T1]{fontenc}
\usepackage{dsfont}
\usepackage{youngtab}
\usepackage{amsfonts}
\usepackage{mathrsfs}
\usepackage{array}
\usepackage{float}
\usepackage{graphicx, rotating}
\usepackage{epstopdf}
\usepackage{epsfig}
\usepackage{cite}
\usepackage{latexsym}
\usepackage{color}
\usepackage{bm}
\usepackage{slashed}
\usepackage{comment}
\usepackage[pagebackref=false]{hyperref}
\definecolor{rossos}{cmyk}{0,1,1,0.55}
\definecolor{bluscuro}{rgb}{0.15, 0.2, .85}
\definecolor{bluchiaro}{cmyk}{1,.3,0.,0.1}
\hypersetup{colorlinks, citecolor=bluscuro, linkcolor=bluscuro, urlcolor=bluscuro}
\def\16{{\bf 16}}
\def\1{{\bf 1}}

\def\2{{\bf 2}}
\def\3{{\bf 3}}
\def\4{{\bf 4}}

\def\be{\begin{equation}}
	\def\ee{\end{equation}}

\def\bar{\overline}

\font\teneurm=eurm10 \font\seveneurm=eurm7 \font\fiveeurm=eurm5
\newfam\eurmfam
\textfont\eurmfam=\teneurm \scriptfont\eurmfam=\seveneurm
\scriptscriptfont\eurmfam=\fiveeurm

\font\teneusm=eusm10 \font\seveneusm=eusm7 \font\fiveeusm=eusm5
\newfam\eusmfam
\textfont\eusmfam=\teneusm \scriptfont\eusmfam=\seveneusm
\scriptscriptfont\eusmfam=\fiveeusm

\font\tencmmib=cmmib10 \skewchar\tencmmib='177
\font\sevencmmib=cmmib7 \skewchar\sevencmmib='177
\font\fivecmmib=cmmib5 \skewchar\fivecmmib='177
\newfam\cmmibfam
\textfont\cmmibfam=\tencmmib \scriptfont\cmmibfam=\sevencmmib
\scriptscriptfont\cmmibfam=\fivecmmib

\numberwithin{equation}{section}

\def\Rn{{R\'{e}nyi }}

\def\be{\begin{equation}}
\def\ee{\end{equation}}
\def\bg{\begin{gather}}
\def\eg{\end{gather}}
\def\ba#1\ea{\begin{align}#1\end{align}}
\def\bg#1\eg{\begin{gather}#1\end{gather}}
\def\bm#1\em{\begin{multline}#1\end{multline}}
\def\bmd#1\emd{\begin{multlined}#1\end{multlined}}

\begin{document}
	\begin{titlepage}
	
	\hspace{11cm}	IFT-UAM/CSIC-21-23

\vskip 1.5in
	
	\begin{center}
		{\bf\Large{Ergodic Equilibration of \Rn Entropies}}\\[.5cm]
		{\bf\Large{and Replica Wormholes}}
		\vskip
		0.6cm {Martin Sasieta} \vskip 0.6cm {\small{ \textit{Instituto de F\'{i}sica Te\'{o}rica, IFT-UAM/CSIC}\\[.15cm]
				{\textit{c/ Nicol\'{a}s Cabrera 13, Universidad Aut\'{o}noma de Madrid, 28049, Madrid, Spain}}}\\[.3cm]
			{\it E-mail}:  \href{mailto:martin.sasieta@uam.es}{\nolinkurl{martin.sasieta@csic.es}}\\
		}
	\end{center}
	\vskip 0.5in
	\baselineskip 16pt
	\begin{abstract}   
		We study the behavior of \Rn entropies for pure states from standard assumptions about chaos in the high-energy spectrum of the Hamiltonian of a many-body quantum system. We compute the exact long-time averages of \Rn entropies and show that the quantum noise around these values is exponentially suppressed in the microcanonical entropy. For delocalized states over the microcanonical band, the long-time average approximately reproduces the equilibration proposal of H. Liu and S. Vardhan, with extra structure arising at the order of non-planar permutations. We analyze the equilibrium approximation for AdS/CFT systems describing black holes in equilibrium in a box. We extend our analysis to the situation of an evaporating black hole, and comment on the possible gravitational description of the new terms in our approximation.
		
	\end{abstract}

\end{titlepage}

{
	\hypersetup{linkcolor=black}
	\tableofcontents
}

\section{Introduction}

Recent attempts to derive the Page curve from semiclassical Euclidean gravity in low-dimensional models have proven remarkably successful \cite{PSSY,AHMST, GHT}. The lesson is that the sharp change in tendency of the Page curve for \Rn entropies can be reproduced \textit{á la} Hawking-Page from an exchange in dominance at the Page time between the disconnected Euclidean black hole saddle and the so-called `replica wormhole' saddle. For the purity of the state of the radiation $\rho_R$, the replica calculation in gravity outputs a formula consisting of these two leading contributions
\be\label{trrhosq}
\text{Tr}_R\,\rho_R^2\; \approx \; e^{-S_{\beta}^R}\,+\,e^{-S_{\beta}^{\bar{R}}}\,\;\;,
\ee
where $S^{R}_{\beta}$ is the thermal second \Rn entropy of the radiation and $S^{\bar{R}}_{\beta}$ is the thermodynamic `Bekenstein-Hawking' second \Rn entropy of the black hole at inverse temperature $\beta$.

Even if replica wormholes `unitarize' the \Rn entropy of the radiation, their inclusion seems to lead to a fundamental incompatibility with a conventional quantum mechanical description. The semiclassical computation of the state of the radiation $(\rho_R)_{ij}$ produces the well-known thermal result, up to perturbative corrections. The most natural way to reconcile a thermal density matrix $(\rho_R)_{\beta}$ with \eqref{trrhosq} is to declare that Euclidean gravity is effectively reproducing an averaged description of the `pseudo-random' properties of the discrete UV spectrum of the theory \cite{Stanford, Liu,Pollack, Altland,Harlow:2020bee,BelindeBoer,  Belin:2020jxr, Chen:2020ojn, Hsin:2020mfa, Krishnan}.

In this note, we reexamine the possibility introduced in \cite{Liu} that replica wormholes give an approximation to the equilibrated  physics of a single unitary theory. We appeal to quantum ergodicity in a high-energy microcanonical band of the Hamiltonian and derive a microscopic version of the `equilibrium approximation' of \cite{Liu} for \Rn entropies. Our results have some additional structure that enters at the level of non-planar diagrams of the previous approximation. We then show that for a certain class of states our results can be further approximated by the microcanonical and the canonical ensembles. In the case of the canonical ensemble, each term of our approximation can be computed from an Euclidean path integral over replicas of the system with different patterns of connectivity between them.

 We then consider initial pure states in AdS/CFT systems that lead to a large black hole in equilibrium with its radiation. For the purity of the radiation subystem, the equilibrium approximation in this case consists of three terms
 \be\label{trrhosqder}
 \left({\text{Tr}_R\,\rho_R^2}\right)_{\text{eq}}\; \approx  \; e^{-S_{\beta}^R}\,+\,e^{-S_{\beta}^{\bar{R}}}\,-\,e^{-(S_{\beta}^R+S_{\beta}^{\bar{R}})}\;\;.
 \ee
 The first two terms agree with \eqref{trrhosq} and, in this case, the equilibrium approximation itself prescribes a boundary path integral to compute each of these terms. The pattern of connectivity between the replicas in each of these path integrals is ultimately related to the topology of the corresponding saddle of the semiclassical gravitational path integral. Similar considerations hold for higher \Rn entropies.
 
 We finally engineer a different AdS/CFT setup that contains an evaporating black hole in the quasi-equilibrium approximation. We get a similar qualitative result to \eqref{trrhosqder} for the purity of the radiation at each epoch of evaporation. The last term in \eqref{trrhosqder} is responsible for recovering the exact pure state for the radiation $\rho_R$ at the endpoint of evaporation. The contribution of these terms is reminiscent of higher genus saddles to the gravitational path integral in models of JT gravity. The minus sign in \eqref{trrhosqder}, however, seems to be prescribed from the exact unitary description and it might be \textit{ad hoc} from semiclassical gravity.

The paper is organized as follows: In section $\ref{sec::2}$ we derive the microscopic equilibrium approximation for \Rn entropies from standard properties of chaotic many-body quantum systems. In section $\ref{sec::3}$ we give an estimation of the average quantum noise around the microscopic equilibrium value. In section $\ref{sec::4}$ we consider a class of delocalized states over a microcanonical band and we obtain the microcanonical and canonical equilibrium approximations for the \Rn entropies. In section $\ref{sec::5}$ we analyze the equilibration of \Rn entropies for a black hole inside a finite box, and obtain \eqref{trrhosqder} and higher \Rn analogs. We also consider the case of a slowly evaporating black hole. We end with some conclusions and appendix \ref{appendix::A} containing some technical details.

\section{Ergodicity and Long-Time Averaging}
\label{sec::2}

In this section, we will study the behavior of \Rn entropies for pure states under mild assumptions about chaos in a high-energy microcanonical band of the Hamiltonian of a many-body quantum system. We will start by considering a microcanonical band $\mathcal{H}_{E,\epsilon}$ of the Hamiltonian $H$ consisting of states with energies in the interval $[E-\epsilon,\, E+\epsilon]$. The energy window will be narrow $\epsilon \ll E$, but still spacious enough to accommodate a large microcanonical entropy $S = \log \mathcal{N}$, where $\mathcal{N}$ is the number of energy eigenstates $\lbrace\ket{E_i}\rbrace$ on the band. We will assume that the spectrum of the Hamiltonian $\lbrace E_i\rbrace$ is non-degenerate in this band\footnote{The non-degeneracy condition is broken whenever the chaotic system has a global symmetry $G$, discrete or continuous. Our analysis directly applies restricting to a superselection sector of $G$.}, and that there are no rational relations between different energy eigenvalues\footnote{For a holographic CFT on $\mathbf{S}^{d-1}\times \mathbf{R}$ of radius $R$, this condition will not be strictly satisfied in the high-energy spectrum $ER \sim \mathcal{O}(c)$ due to rational relations and degeneracies of the descendant states. However, since most of the spectrum will be generated by new primaries, these effects are expected to be of subleading order in $1/\mathcal{N}$ as well as highly dependent on the particular operator content of the theory. A possibility is to slightly break conformal symmetry (e.g. by adding a small relevant deformation or by coupling the CFT to an external system) which will make the spectrum generic. }. In particular, the energy differences $E_i-E_j$ will be rationally independent, which translates into the lack of resonances in the system. 

Given some initial state localized on this band $\ket{\Psi}  = \sum_i\, c_{i}\ket{E_i}$, its time evolution will involve an effective number of $\mathcal{N}_{\text{eff}}$ energy eigenstates, where
\begin{equation}
\mathcal{N}_{\text{eff}}^{-1}\, = \, \sum_i|c_i|^4\;\;.
\end{equation} 
For any $t>0$, the position of the state vector $\ket{\Psi(t)}$ lies on a torus determined by these $\mathcal{N}_{\text{eff}}$ real phases, $\mathbf{T}^{\,\mathcal{N}_{\text{eff}}}$, and in fact the lack of resonances will make it an ergodic cover of this torus. 

In what follows, we will assume a bipartition of the full system, $\mathcal{H}\, =\, \mathcal{H}_R\otimes \mathcal{H}_{\bar{R}}$, and study the time-evolution of the entanglement spectrum of $\ket{\Psi(t)}$ with respect to this bipartition. More precisely, we will study the set of \Rn entropies 
\begin{equation}\label{thermrenyi}
	\mathcal{Z}_n^{(R)}\, = \, e^{-(n-1) S_n^{(R)}}\,= \, \text{Tr}_R\,\rho_R^n\;\;,
\end{equation}
where $\rho_R = \text{Tr}_R \ket{\Psi(t)}\bra{\Psi(t)}$ is the reduced density matrix of subsystem $R$. We will exploit the spectral properties of $H$ on the band to compute the long-time averages of \Rn entropies
\begin{equation}\label{longtimeav}
\overline{\mathcal{Z}^{(R)}_n}\, = \, \lim_{T\rightarrow \infty}\,\dfrac{1}{T}\,\int_{0}^T\,\text{d}t\,\mathcal{Z}^{(R)}_n(t)\;\;,
\end{equation}
which, as we shall see, will tell us an idea about `equilibrium' values of \Rn entropies.

\begin{figure}[h]
	\centering
	\includegraphics[width = .7\textwidth]{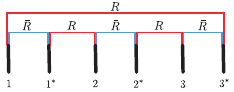}
	\caption{Schematic representation of $\bra{R,\bar{R}\,}$ for $n=3$, where the connections represent contractions of subsystem indices of the corresponding factors.}
	\label{fig::bratraces}
\end{figure}

For later convenience, and in order to make contact with \cite{Liu}, we will introduce some notation. Density matrices like $\ket{\Psi}\bra{\Psi}$ can be viewed as states $\ket{\Psi}\otimes \ket{\Psi}^* \in \mathcal{H} \otimes \mathcal{H}$, where the star denotes some antiunitary operation like CPT. Similarly, $\mathcal{Z}^{(R)}_n$ can be regarded as an amplitude on $(\mathcal{H}\,\otimes \,\mathcal{H})^n$, namely as $\mathcal{Z}^{(R)}_n = \bra{R,\bar{R}\,} \, \left(\ket{\Psi(t)}\otimes \ket{\Psi(t)}^*\right)^n$. All the information about the partial tracing is kept in the bra $\bra{R, \bar{R}\,}$ that lives in the dual space to this replicated Hilbert space (see Fig. \ref{fig::bratraces}). 

We will also define a set of states on the replicated Hilbert space that will turn out to be particularly useful for notational purposes. Given a density matrix $\rho = \sum \rho_{ij}\ket{E_i}\bra{E_j}$, and some permutation $\sigma \in S_n$, we define a state $\ket{\rho,\sigma} \in (\mathcal{H} \otimes \mathcal{H})^n$ as
\be
\ket{\rho,\sigma}\, \equiv \,\sum_{i_k,j_k} \,\rho_{i_1j_{\sigma(1)}}\,...\;\rho_{i_nj_{\sigma(n)}}\,\ket{E_{i_1}, E^*_{j_1},\, ... \,, E_{i_n}, E^*_{j_n}}\;\;.
\ee

First, we will compute the long-time average of the purity, $\mathcal{Z}^{(R)}_2 = \text{Tr}_R \,\rho_R^2$. The long-time integral \eqref{longtimeav} in this case involves four replicas and is given by
\be\label{trrho2}
\left({\mathcal{Z}^{(R)}_2}\right)_{\text{eq}}\,  =   \, \,\sum_{i_1,i_2,j_1,j_2}\,c_{i_1} \,c_{j_1}^* \,c_{i_2} \,c_{j_2}^*\,\overline{e^{-it(E_{i_1}+E_{i_2}-E_{j_1}-E_{j_2})}}\,\braket{R,\bar{R}\,|\, E_{i_1}, E^*_{j_1}, E_{i_2}, E^*_{j_2}}\,\;.
\ee
The long-time integral of the phase vanishes whenever the total frequency is non-zero, which in this case requires $E_{i_1} + E_{i_2} = E_{j_1} + E_{j_2}$. From the assumption that there are no rational relations between energy eigenvalues, this condition can only hold when the $i$ and $j$ eigenvalues are identified, which leads to three different long-time saddles
\be\label{secondm}
\overline{e^{-it(E_{i_1}+E_{i_2}-E_{j_1}-E_{j_2})}}\, = \delta^{i_1}_{j_1}\delta^{i_2}_{j_2}\,+\, \delta^{i_1}_{j_2}\delta^{i_2}_{j_1}\,-\,\delta^{i_1}_{i_2}\delta^{i_1}_{j_1}\delta^{i_2}_{j_2}\;\;,
\ee
and we are not adopting the convention of summing over repeated indices for the last term. This last term is essential in order to avoid over-counting for the case of the configuration $E_{i_1} = E_{i_2} = E_{j_1} = E_{j_2}$ (see Appendix \ref{appendix::A}).

Following these considerations, we derive the average value of the purity
\be\label{secondrenyi}\
\overline{\mathcal{Z}_2^{(R)}}\,   =   \, \text{Tr}_{R}\left(\text{Tr}_{\bar{R}}\,\rho\right)^2\,+\,\text{Tr}_{\bar{R}}\left(\text{Tr}_{R}\,\rho\right)^2\,-\,\braket{R,\bar{R}\,|\,\phi} \;\;,
\ee
for the microscopic equilibration density matrix $\rho = \sum_i\, |c_i|^2\,\ket{E_i}\bra{E_i}$. The unnormalized state $\ket{\phi}$ is a mutipartite entangled state in the replicated Hilbert space 
\be
\ket{\phi} \, = \,\sum_i\, |c_i|^4\, \ket{E_{i},\, E^*_{i},\, E_{i},\, E^*_{i}}\;.
\ee
The first two terms in \eqref{secondrenyi} match the equilibrium proposal of \cite{Liu} but in this case $\rho$ possesses microscopic information about the initial pure state $\ket{\Psi}$. From the symmetries of $\ket{\phi}$ under permutations of the replicas, it is straightforward to see that the whole expression is invariant under $R\leftrightarrow \bar{R}$, which is expected from an average over the long-time ensemble of pure states. 

The new term that we obtain is essential to preserve exact purity in the limit in which $R$ becomes the whole system, $\mathcal{H}_{{R}} = \mathcal{H}$. In this limit, $\text{Tr}_R \,\rho = 1$ and $\text{Tr}_R \,\rho^2 = \mathcal{N}_{\text{eff}}^{-1}$, so the long-time average of the purity becomes
\be
\overline{\mathcal{Z}^{(R)}_2} \, \longrightarrow \; \dfrac{1}{\mathcal{N}_{\text{eff}}}\,+\,1\,-\,\dfrac{1}{\mathcal{N}_{\text{eff}}}\, = \, 1\;\;,
\ee
which is consistent with exact unitarity, from $\overline{\text{Tr}\left(\ket{\Psi(t)}\bra{\Psi(t)}\right)^2} = 1$.

The equilibration value of $\mathcal{Z}_n^{(R)} = \text{Tr}_R\, \rho_R^n$ will similarly be given by a long-time integral over the $2n$ replicas
\be\label{nlongtime}
\overline{\mathcal{Z}_n^{(R)}}\,  =\,  \sum_{i_k,j_k}\,c_{i_1}c_{j_1}^*\,...\,c_{i_n}c_{j_n}^*\,\overline{e^{-it\left(E_{i_1}-E_{j_1}+\,...\,+E_{i_n}-E_{j_n}\right)}}\,\braket{R,\bar{R}\,|\,E_{i_1},E^*_{j_1},\,...\,, E_{i_n}, E^*_{j_n}}\;.
\ee
The long-time integral of the phase again imposes the constraint that the total frequency is zero, $E_{i_1}+...+E_{i_n} = E_{j_1}+...+E_{j_n}$. The lack of rational relations in the spectrum of the Hamiltonian allows to perform this integral without the need to know the particular spectrum, mainly reducing the integral to a simple combinatorial problem for the long-time saddles, which is explained in detail in Appendix \ref{appendix::A}. We import the result here
\begin{equation}\label{nmomentbody}
	\overline{e^{-it\left(E_{i_1}-E_{j_1}+...+E_{i_n}-E_{j_n}\right)}}\,  =   \, n!\;\delta^{(i_1}_{j_1}...\delta^{i_n)}_{j_n}\,\sum_{\pi \in \Pi_n}\,\alpha_{\pi}\,\prod_{B\in \pi}\,\prod_{a,b \in B}\,\delta^{i_{a}}_{i_{b}}\;\;,
\end{equation}
where the sum is over partitions $\Pi_n$ of the set $\mathbf{N}_n = \lbrace 1,2,...,n \rbrace$. A given partition has the form $\pi = \lbrace B_1,...,B_r\rbrace $ and the $B_k$ are `boxes' containing $n_k$ elements of $\mathbf{N}_n$. The number of terms in the sum is the number of partitions of $\mathbf{N}_n$, which is known as the $n$-th Bell number $\mathcal{B}_n$, and grows super-exponentially for large $n$. The coefficients $\alpha_\pi$ can be recursively found from the relation
\begin{equation}
\alpha_{\pi} \, = \,  \dfrac{1}{n_1!\,...\,n_r!}\,-\,\sum_{\pi ' < \pi }\,\alpha_{\pi'}\;\;,
\end{equation}
where $\pi' < \pi$ represents the sum over finer partitions $\pi'$. The coefficient for the finest partition $\pi_e = \lbrace 1,2,...,n\rbrace$ is set to $\alpha_{\pi_e} = 1$ in this normalization. The values of $\alpha_{\pi}$ for some of the finest partitions are explicitly computed in Appendix \ref{appendix::A}.

In this way, we arrive to the long-time average of the $n$-th \Rn entropy
\begin{equation}\label{nrenyi}
\overline{\mathcal{Z}_n^{(R)}}\,  =\,  \sum_{\sigma \in S_n}\,\bra{R,\bar{R}\,}\left.{\rho,\sigma}\right>\;+\; \sum_{\pi \in \Pi_n^*}\,{\alpha}_\pi \,|S_\pi| \,\sum_{\sigma \in S_n/S_\pi }\bra{R,\bar{R}\,}\left.{\phi_{\pi},\sigma}\,\right>\;.
\end{equation}
The first sum exactly reproduces the terms in the equilibrium \textit{ansatz} of \cite{Liu}, but again one has to consider the microscopic equilibration density matrix $\rho = \sum |c_i|^2\ket{E_i}\bra{E_i}$ which has information about the initial state $\ket{\Psi}$ of the system. The second term is a sum over non-trivial partitions $\Pi_n^* = \Pi_n \backslash \lbrace \pi_{e}\rbrace $ and over the permutation orbit $S_n/S_\pi$ of each partition $\pi$. Here $S_\pi < S_n$ the stabilizer subgroup of a given partition $\pi$, which can be intuitively understood as any permutation that preserves the content of the `boxes' $B_l\in \pi$. The new states correspond to different multipartite entangled states 
\begin{equation}\label{newstates}
\ket{\phi_\pi,\, \sigma}\, = \, \, \sum_{i_k}\,\prod_{B\in \pi}\,\prod_{a,b \in B}\,\delta^{i_{a}}_{i_{b}}\,|c_{i_1}|^2\,...\,|c_{i_n}|^2\,\ket{E_{i_1}, E^*_{i_{\sigma(1)}},\,...\,, E_{i_n}, E^*_{i_{\sigma(n)}}}\;\;,
\end{equation}
with a new `source' of entanglement coming from the projections $\delta^{i_a}_{i_b}$ associated to each partition $\pi$. We can also write $\ket{\phi_\pi,\, \sigma}\, = P_\pi \,\ket{\rho,\sigma}$, for $P_\pi$ the projection operator acting as in \eqref{newstates} on half of the $\mathcal{H}$-factors of the replica Hilbert space. The number of new states $\ket{\phi_\pi,\, \sigma}$ also scales super-exponentially with $n$. These states are essential to restore the exact purity in \eqref{nrenyi} when the subsystem $R$ is allowed to gradually approach the size of the full system. 

The long-time average of the \Rn entropy \eqref{nrenyi} is invariant under $R\leftrightarrow \bar{R}$. This property follows from $\bra{R,\bar{R}} \left.{\rho,\sigma}\right>= \bra{\bar{R},R}\left.{\rho,\sigma\tau}\right> $, with the cyclic permutation $\tau = (1,2,...,n-1,n)\in S_n$. The right-multiplication is an isomorphism in $S_n$ and therefore the first sum is trivially invariant. All possible partitions are present for each permutation in the second sum, making the whole sum also invariant.

\begin{figure}[h]
	\centering
	\includegraphics[width = .7\textwidth]{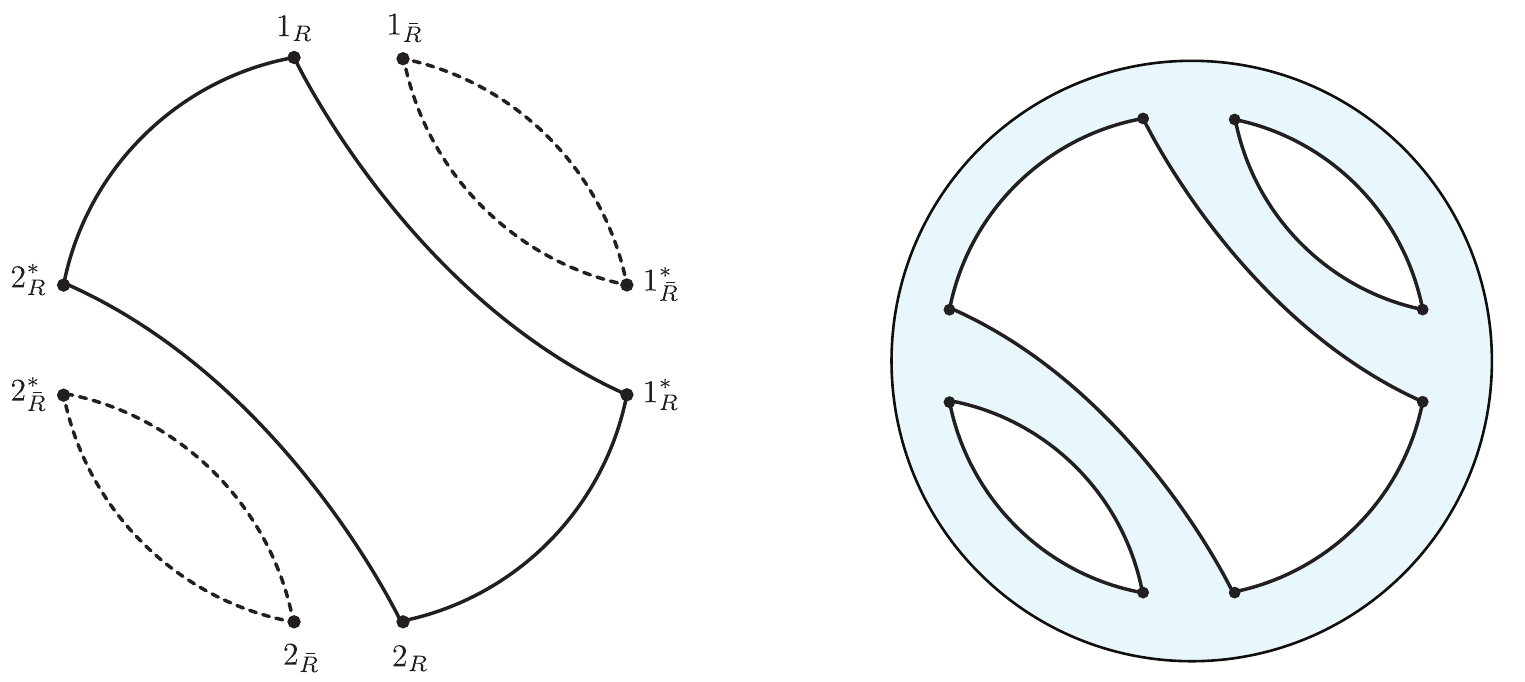}
	\caption{On the left, the diagram corresponding to the amplitude $\bra{R,\bar{R}} \left.{\rho,e}\right>$ for $n=2$. The lines along the circumference are the boundary conditions imposed by the bra, while the lines of the interior are prescribed by the particular permutation, $\sigma = e$ in this case. Each solid loop corresponds to a factor of $n_{R}$ and each dashed loop corresponds to a factor of $n_{\bar{R}}$. On the right, the double-line diagrammatic representation of the amplitude. In this representation, $k(\eta^{-1} \,e)\,+\,k(e) = F-1 = 3$, where $F$ is the number of white faces (with the exterior included). In general, $k(\eta^{-1} \,\sigma)\,+\,k(\sigma) = F-1 = E-V - 2g+1$ from the Euler-Poincaré formula, where $E$ is the number of edges and $V$ is the number of vertices of the blue `polygon', while $g$ is the genus of the surface in which the polygon is embedded. The inequality \eqref{ineq1} follows from $E=3n$ and $V= 2n$.}
	\label{fig::diagramone}
\end{figure}

We can estimate the magnitude of each term in \eqref{nrenyi} if we introduce the effective rank $n_{R}$ and $n_{\bar{R}}$ of the density matrix $\rho$ on each of the subsystems
\begin{gather}
(n_R)^{-1} \equiv \text{Tr}_R\,\left(\text{Tr}_{\bar{R}}\,\rho\right)^2\;\;,\\
(n_{\bar{R}})^{-1} \equiv \text{Tr}_{\bar{R}}\,\left(\text{Tr}_R\,\rho\right)^2\;\;.
\end{gather}
The matrix elements of $\rho$ in an orthonormal basis $\lbrace{\ket{r,\bar{r}} }\rbrace$ of $\mathcal{H}_R\otimes \mathcal{H}_{\bar{R}}$ will have magnitude $(\rho)_{r\bar{r}r'\bar{r}'}\sim (n_Rn_{\bar{R}})^{-1}$. In this basis, each amplitude is
\begin{equation}
\bra{R,\bar{R}\,}\left.{\rho,\sigma}\right>\, = \, \sum_{r_i,\bar{r}_i,r_i',\bar{r}_i'}\,\prod_{i}\,\delta^{r_{i}}_{r'_{\eta(i)}}\,\delta^{\bar{r}_{i}}_{\bar{r}'_{i}}\,\left(\rho\right)_{r_i\bar{r}_ir'_{\sigma(i)}\bar{r}'_{\sigma(i)}}\;\;,
\end{equation}
where $\eta = (n,n-1,...,2,1) \in S_n$. Following the argument in \cite{Liu}, the diagrammatic representation of each amplitude shows the number of $R$ and $\bar{R}$ loops present, and each loop corresponds to a factor of $n_R$ and $n_{\bar{R}}$ respectively  (see Fig \ref{fig::diagramone}). The number of $\bar{R}$ loops in $\bra{R,\bar{R}\,}\left.{\rho,\sigma}\right>$ can be shown to be the number of cycles of the permutation $\sigma$, denoted $k(\sigma)$.  The number of $R$-loops, on the other hand, can be shown to be $k(\eta^{-1} \sigma)$. Altogether, the magnitude of the amplitude is
\begin{equation}\label{permclass}
\bra{R,\bar{R}\,}\left.{\rho,\sigma}\right>\,\sim\, \dfrac{(n_R)^{k(\eta^{-1} \sigma)}\,(n_{\bar{R}})^{k(\sigma)}}{(n_R\,n_{\bar{R}})^n}\;\;.
\end{equation}
The dominant permutations corresponding to planar diagrams are constructed from the `non-crossing partitions' of $\mathbf{N}_n$ and saturate the inequality 
\begin{equation}\label{ineq1}
	 k(\eta^{-1} \sigma)\,+\,k(\sigma)\, \leq \,n+1\;\;.
\end{equation}
For $n_R < n_{\bar{R}}$ the leading permutation in \eqref{permclass} is $\sigma = e$, while for $n_R >\,n_{\bar{R}}$ it is $\eta$.

Let us define the `order' of a given partition $\pi = \lbrace B_1, ..., B_r \rbrace$ of boxes of size $|B_k| = n_k$ as the product $|\pi| \equiv n_1\,n_2\,...\,n_k$. Then, we can see that the effect of the projector $P_\pi$ in $\bra{R,\bar{R}\,}\left.{\phi_{\pi},\sigma}\right>$  is nothing but to reduce the previous result by a factor of $\mathcal{N}_{\text{eff}}^{1-|\pi|}$. The contribution of the new states is then
\begin{equation}\label{projclass}
\bra{R,\bar{R}\,}P_\pi\,|\left.{\rho,\sigma}\right>\,\sim\, \dfrac{(n_R)^{k(\tau \sigma)}\,(n_{\bar{R}})^{k( \sigma)}}{(n_R\,n_{\bar{R}})^{n}\,\mathcal{N}_{\text{eff}}^{|\pi|-1}}\;\;,
\end{equation}
and in particular we can see that $|\pi| >1$ implies that the new terms enter the long-time value at least at the order of permutations with non-planar diagrams. Note, however, that this hierarchy is less pronounced when $\ket{\Psi}$ involves a small number of energy eigenstates, and in this case the new terms can become comparable to certain `planar' permutations. In fact, for a single eigenstate, all the terms in \eqref{nrenyi} are of the same order of magnitude.

\section{Quantum Noise}
\label{sec::3}

In the previous section, we have shown that the long-time average of the \Rn entropies in a microcanonical band of a many-body chaotic system produces a microscopic version of the equilibrium \textit{ansatz} proposed in \cite{Liu}. In this section, we will show that when $n_R$ and $n_{\bar{R}}$ are large, quantum fluctuations are suppressed with respect to the average value in the long run. In this sense, it is reasonable to expect that the long-time average is a measure of the `equilibrated' value of the \Rn entropy, at least for timescales $t \ll t_P \sim E^{-1}\exp(\mathcal{N}_{\text{eff}})$ with no Poincaré recurrences on the system.

In order to simplify the discussion, we will neglect the terms coming from the projectors since they will be always subdominant with respect to leading permutations. The long-time variance $\Delta^{(R)}_n$ of the \Rn entropy is defined as the square root of
\begin{equation}
		\overline{\left(\mathcal{Z}^{(R)}_n\,-\,\overline{\mathcal{Z}^{(R)}_n}\right)^2}\, \approx \, \sum_{\sigma \in S_{2n}}\,\left(\bra{R,\bar{R}}\otimes \bra{R,\bar{R}\,}\right)\,|\left.\rho,\sigma \right\rangle\;-\;\sum_{\sigma,\sigma' \in S_{n}}\,\bra{R,\bar{R}}\left.\rho,\sigma \right\rangle\,\bra{R,\bar{R}}\left.\rho,\sigma' \right\rangle\;\;,
\end{equation}
which can be written in the compact notation
\begin{equation}\label{ltvariance}
 \left(\Delta^{(R)}_n\right)^2\, \approx \, \sum_{\sigma \in A_{2n}}\,\left(\bra{R,\bar{R}}\otimes \bra{R,\bar{R}\,}\right)\,|\left.\rho,\sigma \right\rangle\;\;,
\end{equation}
where $A_{2n} = S_{2n} \,\backslash\, S_{n}\times S_n$ is the set of `connected' permutations between the two $\mathcal{Z}^{(R)}_n$ factors.

\begin{figure}[h]
	\centering
	\includegraphics[width = .35\textwidth]{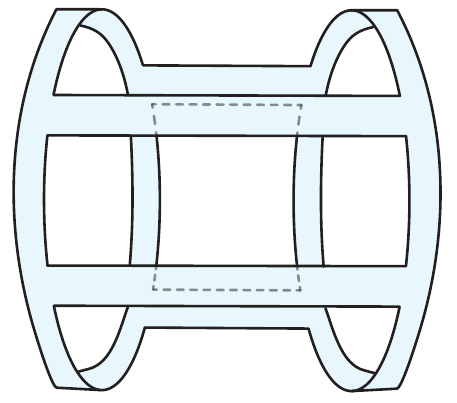}
	\caption{Double-line diagram corresponding to $\left(\bra{R,\bar{R}}\otimes \bra{R,\bar{R}\,}\right)\,|\left.{\rho,\sigma}\right>$ for $n=2$ and the connected permutation $\sigma = (13)(24) \in A_4$. For connected diagrams, the total number of loops is $k(\tau_2 \sigma) + k(\sigma) = F-2 = E-V-2g$, with $E=6n$ and $V= 4n$. Therefore, only planar permutations $\sigma \in A_n$ saturate \eqref{cota}. }
	\label{fig::diagramtwo}
\end{figure}

The magnitude of each term in \eqref{ltvariance} can also be estimated from a double-line diagrammatic counting \textit{à la} 't Hooft (see Fig. \ref{fig::diagramtwo}). In the product basis, $\lbrace{\ket{r,\bar{r}} }\rbrace$, each amplitude is
\begin{equation}
	\left(\bra{R,\bar{R}}\otimes \bra{R,\bar{R}\,}\right)\,|\left.{\rho,\sigma}\right>\, = \, \sum_{r_i,\bar{r}_i,r_i',\bar{r}_i'}\,\prod_{i}\,\delta^{r_{i}}_{r'_{\eta_2(i)}}\,\delta^{\bar{r}_{i}}_{\bar{r}'_{i}}\,\left(\rho\right)_{r_i\bar{r}_ir'_{\sigma(i)}\bar{r}'_{\sigma(i)}}\;\;,
\end{equation}
where $\eta_2 = (2n,...,n+1)(n,...,1) \in S_{2n}$. The number of $\bar{R}$-loops of the diagram does not change with respect to the previous estimation, since $\bra{R,\bar{R}}_n\otimes \bra{R,\bar{R}}_n$ has the same $\bar{R}$-tracing pattern as the bra $\bra{R,\bar{R}}_{2n}$, where the subindex represents the number of replicas. Therefore, we will have $k(\sigma)$ of such loops, each of them yielding a factor of $n_{\bar{R}}$. The number of $R$-loops, however, notices the new `factorized' tracing pattern and the total number will be given in this case by $k(\tau_2 \sigma)$, where $\tau_2 = \eta_2^{-1}\, =  (1,...,n)(n+1,...,2n) \in S_{2n}$. The total contribution is then
\begin{equation}\label{varclass}
\left(\bra{R,\bar{R}}\otimes \bra{R,\bar{R}\,}\right)\,|\left.{\rho,\sigma}\right>\,\sim\, \dfrac{(n_R)^{k(\tau_2 \sigma)}\,(n_{\bar{R}})^{k(\sigma)}}{(n_R\,n_{\bar{R}})^{2n}}\;\;,
\end{equation}
Note that for any `disconnected' $\sigma \in S_n \times S_n$ this expression recovers the square of \eqref{permclass}. However, disconnected permutations are not allowed because we need to restrict to $A_{2n}$. In particular, we have that for $\sigma \in A_{2n}$ the following inequality holds
\begin{equation}\label{cota}
	k(\tau_2 \sigma)\, + \, k(\sigma)  \leq 2n\;\;.
\end{equation}

Assuming that $n_R < n_{\bar{R}}$, \eqref{varclass} and \eqref{cota} show that the quantum noise for the \Rn entropy is suppressed by
\begin{equation}
\dfrac{\Delta^{(R)}_n\,}{\overline{\mathcal{Z}^{(R)}_n}}\, \sim \,\dfrac{1}{n_R}\;\;.
\end{equation}
We emphasize that, for finite dimensional systems, $n_R \sim d_R$ is expected to generally be the dimensionality of the subsystem, which scales with the total entropy as $e^{f_RS}$, given a fraction $0<f_R<1/2$ (here we are assuming that $S = \log \mathcal{N}$ accounts for almost the total dimensionality of the system). Therefore, the quantum noise is expected to be suppressed by $\exp (-f_R\,S)$.

In a similar way, we can generalize our results to the $m$-th long-time moment
\begin{equation}
	\overline{\left(\mathcal{Z}^{(R)}_n\,-\,\overline{\mathcal{Z}^{(R)}_n}\right)^m}\, \approx \, \sum_{\sigma \in A_{mn}}\,\left(\bra{R,\bar{R}}^{\otimes m}\right)\,|\left.\rho,\sigma \right\rangle\;\;,
\end{equation}
where $A_{mn} \subset S_{mn}$ consists of totally connected permutations between all of the $\mathcal{Z}^{(R)}_n$ factors. Again, the $\bar{R}$-loops do not notice the different tracing pattern, while the $R$-loops do. The estimation is then
\begin{equation}\label{mvarclass}
	\left(\bra{R,\bar{R}}^{\otimes m}\right)\,|\left.{\rho,\sigma}\right>\,\sim\, \dfrac{(n_R)^{k(\tau_m \sigma)}\,(n_{\bar{R}})^{k(\sigma)}}{(\mathcal{N}_{\text{eff}})^{2n}}\;\;,
\end{equation}
where $\tau_m = (1,...,n)...(n(m-1),...,nm) \in S_{nm}$. In this case, any totally connected permutation $\sigma \in A_{mn}$ will satisfy $	k(\tau_m\sigma)\, + \, k(\sigma)  \leq mn-(m-2)$.

These considerations lead to the conclusion that the long-time averaging induces an effective probability distribution $\mathcal{P}(\mathcal{Z}_n^{(R)})$ for the value of the \Rn entropy which is extremely peaked at the average value \eqref{nrenyi}, with a variance \eqref{ltvariance} which is exponentially suppressed in the effective number of degrees of freedom $\log n_R \gg 1$. The typical timescale for quantum fluctuations is the Heisenberg time $t_H \sim (\Delta E)^{-1}$, where $\Delta E$ is the average energy difference between the eigenstates participating in $\ket{\Psi}$. We expect that in general these fluctuations are effectively `frozen' for timescales $t\ll t_H$ and that the \Rn entropy is `equilibrated' with a value given by \eqref{nrenyi}.

\section{Equilibrium Approximation for \Rn Entropies}
\label{sec::4}

As we have seen, \Rn entropies are very fine-grained measures of the system that retain the information about the initial state $\ket{\Psi}$ for arbitrarily long times. In this section, we will make further assumptions about the Hamiltonian $H$ to approximate the long-time values of \Rn entropies for a general class of initial states by the corresponding `equilibrium values' arising from different thermodynamic ensembles.

To start, we can consider the reduced set of initial states for which the energy wavefunction $c_i$ is delocalized enough such that it excites a large fraction of the energy eigenstates of the band. More precisely, let $\mathcal{N}_{\text{eff}}/\mathcal{N} \,= \,1\,-\,x^2 $ for some $x \ll 1$. For such states, the microscopic equilibration density matrix $\rho$ will be very close to the microcanonical density matrix on the band with respect to the trace distance, $||\rho-\rho_{\text{mc}}||_1 \,\lesssim \, x$. In this case, it is obvious that the long-time purity \eqref{secondrenyi} will be given at leading order in $x$ by
\begin{equation}\label{eqpuritysparse}
\overline{\mathcal{Z}_2^{(R)}}\,\approx \,\left(\mathcal{Z}^{(R)}_2\right)_{\text{eq}}\,   = \, \text{Tr}_R\left(\text{Tr}_{\bar{R}}\,\rho_{\text{mc}}\right)^2\,+\,\text{Tr}_{\bar{R}}\left(\text{Tr}_{{R}}\,\rho_{\text{mc}}\right)^2\,-\, \dfrac{1}{\mathcal{N}}\,\left\langle \mathcal{Z}^{(R)}_{2}\right\rangle \;\;,
\end{equation}
where $\rho_{\text{mc}}$ is the microcanonical density matrix on the band, and 
\begin{equation}
\left\langle \mathcal{Z}^{(R)}_{2}\right\rangle  \, = \, \mathcal{N}^{-1}\,\sum_i \,\text{Tr}_R (\text{Tr}_{\bar{R}}\, \ket{E_i}\bra{E_i})^2\;\;,
\end{equation}
 is the average purity of the energy eigenstates of the band. The first two terms in the expression \eqref{eqpuritysparse} exactly match the microcanonical equilibrium approximation of \cite{Liu}, while the extra term in our approximation comes from the exact unitary description and it is subleading by a factor of $\mathcal{N}^{-1}$.

The requirement $x \ll 1$ is too restrictive. In order to make more general statements corresponding to a much larger set of initial states, we will need to make an extra assumption about the structure of the Hamiltonian. A particularly convenient guiding principle to study quantum chaos is to look at properties of eigenstates of typical Hamiltonians, and to see which of these properties could be approximate features of the chaotic Hamiltonian $H$. One such property is the typicality of the chaotic eigenstates with respect to `small' observables, which is the essence of the \textit{eigenstate thermalization hypothesis} (ETH) \cite{Deutsch,Srednicki1,Srednicki2}.

Our assumption about $H$ will be a lot milder, since we are not going to consider the properties of single eigenstates\footnote{An alternative approach would be to study the equilibration value of the \Rn entropy for initial states involving a few eigenstates, possibly by assuming some property about the chaotic eigenstates like \cite{Dymarsky:2016ntg} or by considering typical eigenstates \cite{Lu:2017tbo, Fujita:2017pju,Srednicki3}.}, but rather averaged properties over a large number of them. Given a state $\ket{\Psi}$ involving a large number of eigenstates $\mathcal{N}_{\text{eff}}$, consider the density matrix $\rho_0 \, = \,\Pi_{\ket{\Psi}} \,\rho_{\text{mc}}\,\Pi_{\ket{\Psi}}$ that is constructed by projecting the microcanonical density matrix $\rho_{\text{mc}}$ into the $\mathcal{N}_{\text{eff}}$-dimensional subspace generated by the eigenstates in which $\ket{\Psi}$ has larger support. Of course, $\rho_0$ is a really good approximation of the microscopic density matrix
\begin{equation}
||\rho - \rho_0 ||_1\,\leq \,\sqrt{\mathcal{N}\,\text{Tr}\left(\rho - \rho_0\right)^2}\,\leq\,\dfrac{2x}{\mathcal{N}_{\text{eff}}}\;\;.
\end{equation}
Note the extra factor of $\mathcal{N}_{\text{eff}}^{-1}$ coming from the fact that $\rho_0$ is a much better approximation to the state $\rho$ than $\rho_{\text{mc}}$.

Let $S_{\text{eff}} = \log \mathcal{N}_{\text{eff}}$ be the `microcanonical entropy' of $\rho_0$. We expect that, whenever this entropy is comparable to the entropy of the band, $(S-S_{\text{eff}})/S \ll 1$, then generally the subset of eigenstates taking part in $\rho_0$ will be a good representative of the full microcanonical ensemble for any quantity which has desirable `convergence' properties, and in particular for the entanglement spectrum of $R$ and $\bar{R}$. Under this assumption, initial states involving a large fraction of the entropy of the band will reproduce the microcanonical value for the long-time average of the \Rn entropy
\begin{equation}\label{mcav}
\overline{\mathcal{Z}_n^{(R)}}\,\approx\,\left(\mathcal{Z}^{(R)}_n\right)_{\text{eq}}\,   =  \, \sum_{\sigma \in S_n}\,\bra{R,\bar{R}\,}\left.{\rho_{\text{mc}},\sigma}\right>\;\;+\; \sum_{\pi \in \Pi_n^*}\,{\alpha}_\pi \,|S_\pi| \,\sum_{\sigma \in S_n/S_\pi }\bra{R,\bar{R}\,} P_\pi| \left.{\rho_{\text{mc}},\sigma}\,\right>\;\;.
\end{equation}

A similar approximation can be done in terms of the canonical ensemble. The thermodynamic entropy can be smoothly defined by $e^{S(E)}\, \equiv \, E\sum_i \delta_{\epsilon}(E-E_i)$, where $\delta_{\epsilon}$ is the `regularized Dirac delta' of width $\epsilon$ that accounts for the discreteness of the spectrum. For the inverse temperature $\beta = \partial S/\partial E$ evaluated at the energy of the band, the ensemble trace distance can be evaluated from a saddle point approximation at large $S$ and yields the well-known result
\begin{equation}
	||\rho_{\text{mc}}-\rho_\beta||_1 \,\sim\, \mathcal{O}(S^{-1})\;\;,
\end{equation}
where $\rho_{\beta} = e^{-\beta H} /Z_\beta $ is the canonical density matrix and $Z_\beta = \text{Tr} \,e^{-\beta H}$ is the canonical partition function. 

For delocalized initial states $\ket{\Psi}$ that excite a large number of eigenstates of the band, we can also approximate their long-time average purity \eqref{secondrenyi} by the canonical equilibration value
\begin{equation}\label{mceq2c}
\overline{\mathcal{Z}^{(R)}_2}\, \approx \, \left(\mathcal{Z}^{(R)}_2\right)_{\text{eq},\,\beta}\,   = \, \text{Tr}_R\left(\text{Tr}_{\bar{R}}\,\rho_{\beta}\right)^2\,+\,\text{Tr}_{\bar{R}}\left(\text{Tr}_{{R}}\,\rho_{\beta}\right)^2\,-\, \dfrac{1}{{Z_\beta}}\,\mathcal{Z}^{(R)}_{2,\beta}\;\;,
\end{equation}
where 
\begin{equation}\label{extrac}
\mathcal{Z}^{(R)}_{2,\beta} \, = \, \left(Z_\beta\right)^{-1}\,\sum_i \,e ^{-2\beta E_i}\,\text{Tr}_R (\text{Tr}_{\bar{R}}\, \ket{E_i}\bra{E_i})^2\;\;.
\end{equation}
The sums are now implicitly ranging over all of the eigenstates of the Hamiltonian. In a completely analogous way, from \eqref{nrenyi} we can obtain the canonical approximation for higher \Rn entropies 
\begin{equation}\label{mceqnc}
\overline{\mathcal{Z}^{(R)}_2}\, \approx \,\left(\mathcal{Z}^{(R)}_n\right)_{\text{eq},\,\beta}\,   = \, \sum_{\sigma \in S_n}\,\bra{R,\bar{R}\,}\left.{\rho_\beta,\sigma}\right>\;\;+\; \sum_{\pi \in \Pi_n^*}\,{\alpha}_\pi \,|S_\pi| \,\sum_{\sigma \in S_n/S_\pi }\bra{R,\bar{R}\,} P_\pi| \left.{\rho_{\beta},\sigma}\,\right>\;\;.
\end{equation}

To further elucidate the structure of the equilibrium approximation for canonical equilibration, let us reintroduce the basis $\lbrace{\ket{r, \bar{r}}\rbrace}$ of $\mathcal{H}_R\otimes \mathcal{H}_{\bar{R}} $. We can write down each term in  \eqref{mceqnc} in this basis in the compact notation
\begin{gather}
\bra{R,\bar{R}\,}\left.{\rho_\beta,\sigma}\right>\, = \, \sum_{r_i,\bar{r}_i,r'_i,\bar{r}'_i}\,\prod_{i}\,\delta^{r_{i}}_{r'_{\eta(i)}}\,\delta^{\bar{r}_{i}}_{\bar{r}'_{i}}\,\braket{r_i, \bar{r}_i\,|\,\rho_{\beta}\,|\,r'_{\sigma(i)}, \bar{r}'_{\sigma(i)}} \;\;,\label{exp1}\\
\bra{R,\bar{R}\,}P_\pi\,|\,\left.{\rho_\beta,\sigma}\right>\, = \, \sum_{r_i,\bar{r}_i,r'_i,\bar{r}'_i}\,\prod_{i}\,\prod_{B\in \pi}\,\prod_{a,b \in B}\,\delta^{r_{c}}_{r_{d}}\,\delta^{\bar{r}_{c}}_{\bar{r}_{d}}\,\delta^{r_{i}}_{r'_{\eta(i)}}\,\delta^{\bar{r}_{i}}_{\bar{r}'_{i}}\, \braket{r_i, \bar{r}_i\,|\,\rho_{\beta}\,|\,r'_{\sigma(i)}, \bar{r}'_{\sigma(i)}}\label{exp2}\;\;.
\end{gather}  

For local theories, each of the terms in terms in \eqref{exp1} and \eqref{exp2} can be written as an Euclidean path integral over $n$ replicas of the system. The amplitudes correspond to the Euclidean time-evolution by an amount $\beta$, with two different states inserted at the boundaries of the strip, $\tau = 0$ and $\tau = \beta$. The delta functions determine the gluing pattern of these $n$ path integrals. To be more specific, let $\Phi_i(\tau) \equiv \lbrace \Phi_i^\mu(\tau,\boldsymbol{x}) \rbrace$ denote the collective set of fields on the $i$-th replica, and let $S_E[\Phi(\tau)]$ be the Euclidean action of the theory. In this notation we have
\begin{gather}
Z_\beta^n\bra{R,\bar{R}\,}\left.{\rho_\beta,\sigma}\right>\, =  \,\int\,\prod_{i=1}^n\,\mathcal{D}\Psi_i \,\mathcal{D}\tilde{\Psi}_i\,\delta(\Psi_i|_R-\tilde{\Psi}_i|_R)\delta(\Psi_i|_{\bar{R}}-\tilde{\Psi}_i|_{\bar{R}})\,\int_{\Phi_i(0) =\tilde{\Psi}_{\sigma(i)} }^{\Phi_i(\beta) = {\Psi}_i} \mathcal{D}\Phi_i(\tau) \,e^{-S_E[\Phi(\tau)]}\;\;,\label{PI1}
\end{gather}
and
\begin{gather}
Z_\beta^n\bra{R,\bar{R}\,}P_\pi|\left.{\rho_\beta,\sigma}\right>\, =  \,\int\,\prod_{i=1}^n\,\prod_{B\in \pi}\,\prod_{a,b \in B}\,\mathcal{D}\Psi_i\, \,\mathcal{D}\tilde{\Psi}_i\,\delta(\Psi_i|_R-\tilde{\Psi}_i|_R)\delta(\Psi_i|_{\bar{R}}-\tilde{\Psi}_i|_{\bar{R}})\,\delta(\Psi_a|_R-{\Psi}_b|_R)\,\times\,\nonumber \\ \times\,\delta(\tilde{\Psi}_a|_{\bar{R}}-\tilde{\Psi}_b|_{\bar{R}})\int_{\Phi_i(0) =\tilde{\Psi}_{\sigma(i)} }^{\Phi_i(\beta) = {\Psi}_i} \mathcal{D}\Phi_i(\tau) \,e^{-S_E[\Phi(\tau)]}\;.\label{PI2}
\end{gather}

We have shown that for a general class of initial states, the long-time averaged values of \Rn entropies will yield a version of the equilibrium \textit{ansatze} of \cite{Liu} for the microcanonical and canonical ensembles, with extra terms that contribute at the level of the non-planar permutations. In our derivation, we mainly used quantum ergodicity of the Hamiltonian $H$, and a restriction to initial states involving a large number of eigenstates of the microcanonical band. Similar results can also be obtained by Haar averaging $\mathcal{Z}_n^{(R)}$ either over initial states $\ket{\Psi}$ in the microcanonical band or over time-evolution operators. Our derivation, on the other hand, directly applies to atypical initial states that remain atypical for $t< t_H$.

\section{Black Hole in a Box and Replica Wormholes}
\label{sec::5}

So far, we have been quite general in our discussion about the nature of the chaotic many-body system under consideration. In this section, we will describe the relevance of the `equilibrium approximation' for \Rn entropies in the context of AdS/CFT systems describing black holes in equilibrium.

We consider a system consisting of a holographic CFT on a spatial sphere $\mathbf{S}^{d-1}$ of radius $\ell_{\text{AdS}}$, which we denote $\mathcal{H}_{\bar{R}}$ . The CFT sphere is contained in an external `radiation box' $\mathcal{H}_{R}$ with no dynamical gravity, and of finite volume $L^{d}$, with $L > \ell_{\text{AdS}}$. The full Hamiltonian of the system is
\begin{equation}\label{Ham}
	H = H_{R}\,+\,H_{\bar{R}}\,+\, H_{\text{int}}\;\;,
\end{equation}
where $H_{R}$ is the weakly coupled Hamiltonian of the box, $H_{\bar{R}}$ is the CFT Hamiltonian, and $H_{\text{int}}$ is a small interaction that allows for transparent boundary conditions in the gravitational description of the system. The Hamiltonian \eqref{Ham} will satisfy the spectral requirements introduced in section \ref{sec::2}, which are mainly inherited from the properties of the black hole band of the CFT Hamiltonian.

The system is initialized at a state $\ket{\Psi} = \sum_i c_i \ket{E_i}$ that belongs to a high-energy microcanonical band of total energy $E \gg L^{-1}$. For our purposes we take $\ket{\Psi}$ to be a semiclassical state of this band such as, for instance, some configuration of matter in AdS. We will assume that the state $\ket{\Psi}$ involves a large number $\mathcal{N}_{\text{eff}}$ of eigenstates of the band. Under time-evolution, the matter will eventually collapse and form a large black hole in AdS. The size of this black hole will depend on the size of the radiation box $L$, where we are assuming that the energy is sufficiently large compared to $L^{-1}$, i.e. $EL \gtrsim (L/l_P)^{\alpha}$ for some $\alpha >1$ that depends on the dimension. 

Strictly speaking, this black hole is not an equilibrium state, since the state vector $\ket{\Psi(t)}$ will indefinitely explore the $\mathcal{N}_{\text{eff}}$-dimensional ergodic torus, $\mathbf{T}^{\mathcal{N}_{\text{eff}}}$. In particular, there will be quasi-periodic quantum fluctuations entering at Heisenberg timescales $t_H \sim (\Delta E)^{-1}$, where $\Delta E$ is the average energy difference between the energy eigenstates participating in $\ket{\Psi}$. In the very long run, at timescales $t_P \sim E^{-1} \exp \mathcal{N}_{\text{eff}}$, these fluctuations will coherently lead to Poincaré recurrences. For certain decaying correlation functions, the quantum noise becomes the leading contribution at late times, and thus the Lorentzian semiclassical description of the state is not good enough to reproduce this non-perturbative quantum gravitational effect \cite{Maldacena:2001kr,Dyson:2002nt,Barbon:2003aq,Barbon:2004ce,Barbon:2014rma, Saad:2018bqo, Saad:2019pqd, Cotler:2020ugk}.

The required time $t_{\text{eq}}$ for the effective equilibration of the entanglement spectrum of $\rho_R$ is expected to be really small compared to $t_H$. For the local Hamiltonian $H_{\bar{R}}$, entanglement is expected to propagate in the form of a wavefront at an effective lightcone velocity $v_{\text{eff}}$  \cite{CalabreseCardy,  Esperanza,  Liu:2013iza,Liu:2013qca, Asplund:2015eha,Casini:2015zua, hartman2015speed, Mezei:2016wfz} and thus timescale for equilibration for the degrees of freedom of the box is of the order $t_{\text{eq}} \sim L/v_{\text{eff}}$. For the black hole degrees of freedom, information spreading occurs much faster, so we expect that the entanglement of these degrees of freedom equilibrates after a few scrambling times $t_{\text{eq}} \sim t_{s}$.
 
For $t\gtrsim t_{\text{eq}}$ we have provided general arguments in the previous sections to declare that the \Rn entropies of the radiation will equilibrate to values which are approximately given by \eqref{mceqnc}, that is
\begin{equation}\label{eqaprox}
\text{Tr}_R\, \rho_R^n (t)\; \approx\; \overline{\text{Tr}_R\, \rho_R^n}\;\approx \; \left(\text{Tr}_R\, \rho_R^n\right)_{\text{eq},\, \beta}\;\;,
\end{equation}
where $\beta = \partial S/\partial E$ is the inverse temperature associated to the microcanonical band. Assuming that the interaction $H_{\text{int}}$ is small, we can neglect its contribution to the canonical ensemble and perform the factorization 
\begin{equation}\label{candm}
\rho_{\beta}\,= \,  \dfrac{e^{-\beta H}}{Z_\beta}\, \approx \, \dfrac{e^{-\beta H_R}}{Z_\beta^R}\,\otimes\, \dfrac{e^{-\beta H_{\bar{R}}}}{Z^{\bar{R}}_{\beta}}\;\;.
\end{equation}

For notational purposes, we will defining the following quantities
\begin{gather}
	Z^R_{n\beta} = \text{Tr}_R\, \left(e^{-\beta H_R}\right)^n\;\;,\\
	Z^{\bar{R}}_{n\beta} = \text{Tr}_{\bar{R}}\,\left(e^{-\beta H_{\bar{R}}}\right)^n\;\;,
\end{gather}
which are in fact related to the thermal \Rn entropies on each of the subsystems.

For the factorized equilibration density matrix \eqref{candm} the canonical equilibration value of the purity \eqref{mceq2c} simplifies and yields
\begin{equation}\label{purityfact}
\left({\text{Tr}_R\,\rho_{R}^2}\right)_{\text{eq}} \;{\approx}\;\;\dfrac{Z^{R}_{2\beta}}{(Z^{R}_{\beta})^2}\,+\,\dfrac{Z^{\bar{R}}_{2\beta}}{(Z^{\bar{R}}_{\beta})^2}\,-\,\dfrac{Z^{R}_{2\beta}}{(Z^{R}_{\beta})^2}\,\dfrac{Z^{\bar{R}}_{2\beta}}{(Z^{\bar{R}}_{\beta_k})^2} \,\;\;,
\end{equation}
Note that the first two terms correspond to  $e^{-S^R_\beta}$ and $e^{-S^{\overline{R}}_\beta}$ from the definition of the thermal second \Rn entropy $S^R_\beta$ and $S^{\overline{R}}_\beta$ of each of the subsystems. The equilibrated value of the purity in terms of the second \Rn entropies is then
\begin{equation}\label{purityfact2}
	\left({\text{Tr}_R\,\rho_{R}^2}\right)_{\text{eq}} \;{\approx}\;\;e^{-S^R_\beta}\,+\,e^{-S^{\overline{R}}_\beta}\,-\,e^{-(S^R_\beta+S^{\overline{R}}_\beta)}\;\;.
\end{equation}

We will now analyze the origin of each term in \eqref{purityfact} from the point of view of the CFT ($\bar{R}$ system). First of all, there is an overall normalization of this expression which is given by $(Z_\beta)^2 \approx (Z_\beta^R)^2\,(Z_\beta^{\bar{R}})^2$ due to the form of the canonical density matrix \eqref{candm}. The numerator of the first term involves a CFT path integral $(Z^{\bar{R}}_\beta)^2$ which precisely cancels the $\bar{R}-$part of this normalization. Therefore, this term corresponds to two disconnected CFT path integrals which in bulk variables will be dominated by a disconnected saddle consisting of two copies of the Euclidean black hole at inverse temperature $\beta$. This disconnected term matches the contribution of the `disconnected saddle' in previous replica calculations in gravity \footnote{In the model of \cite{PSSY}, the radiation Hamiltonian $H_R$ is considered to be a projector into $k$ states of the radiation, which gives $Z^R_{n\beta} = e^{-n\beta}\,k$ or equivalently $S_\beta^{R} = -\log k$. }.

On the other hand, the second term corresponds to the CFT path integral $Z^{\bar{R}}_{2\beta}$ and corresponds to two Euclidean strips of length $\beta$, interpreted in this context as two copies of the CFT system, which are glued together in such a way that they form a single thermal circle of length $2\beta$. Since the replicas are already connected through the boundary conditions, in bulk terms this path integral will be dominated by an Euclidean black hole at inverse temperature $2\beta$ which will connect the two replicas \footnote{We assume that the temperature of the system is well above the Hawking-Page temperature $T\gg R^{-1}$ so that the dominant saddle in $Z^{\bar{R}}_{2\beta}$ is still a black hole.}. This connected term matches the contribution of the `replica wormhole' in previous replica calculations in gravity.

These observations lead to the hypothesis introduced in \cite{Liu} that replica computations of the purity of the radiation using the gravitational path integral $(\text{Tr}_R\,\rho_{R}^2)_{\text{grav}}$ are effectively reproducing each term of the equilibrium approximation \eqref{purityfact2}.

However, the interpretation of the third term in \eqref{purityfact} and \eqref{purityfact2} as arising from a subleading gravitational saddle is less clear. In fact, we can rewrite \eqref{purityfact2} as
\be\label{two}
\left({\text{Tr}_R\,\rho_{R}^2}\right)_{\text{eq}} \;{\approx}\;\; \left(e^{-S^R_\beta}\,+\,e^{-S^{\bar{R}}_\beta}\right)\,\left(1\,-e^{-S^{\bar{R}}_\beta}\,\sum_{m=0}^\infty \,(-1)^m\,e^{-m\left(S^{\bar{R}}_\beta-S^R_\beta\right)}\right)\;\;.
\ee
In JT gravity models \cite{PSSY,AHMST,GHT} the suppression in powers of $S^{\bar{R}}_\beta$ agrees heuristically with the genus expansion in powers of the extremal entropy $S_0$, so these terms might appear at the level of higher genus saddles of the gravitational path integral. However, there is a somewhat obscure minus sign for each handle arising from the global minus sign of the last term in \eqref{purityfact2}. A possibility is that this term is prescribed from the long-time average in the exact unitary description, and that it goes beyond the semiclassical gravitational path integral. In this sense, it can be viewed as a `counterterm' to restore exact unitarity at the level of subleading saddles of the previous two quantities, $Z^{\bar{R}}_{2\beta}$ and $(Z^{\bar{R}}_{\beta})^2$.

Generalizing these results to higher \Rn entropies is again a matter of combinatorics. Let $|\sigma| \equiv k(\sigma)$ denote the number of cycles of $\sigma \in S_n$, each of length $\lbrace s_1,...,s_{|\sigma|}\rbrace$, and similarly for $\sigma' = \tau\sigma$ for the cycle lengths $\lbrace s'_1,...,s'_{|\sigma'|}\rbrace$, where $\tau = (1,...,n)$. Given a non-trivial partition $\pi \in \Pi_n \backslash \lbrace \pi_{e}\rbrace$ and a permutation $\sigma \in S_n$, let $\sigma_\pi \in S_n$ be the `coarse-grained' permutation that is constructed from $\sigma$ by the rule of merging two of its cycles $(a_1,...,a_{L_s})$ and $(b_1,...,b_{L_m})$ whenever some $a$ and some $b$ belong to the same `box' $B_l\in \pi $. Let $|\sigma_\pi| < |\sigma|$ denote the number of cycles of this permutation, of length $\lbrace q_1,...,q_{|\sigma_\pi|}\rbrace$, and similarly for $\sigma'_\pi$, for the lengths $\lbrace q'_1,...,q'_{|\sigma'_\pi|}\rbrace$. In this notation, we can compute \eqref{mceqnc} for the product density matrix \eqref{candm}, which gives
\begin{equation}\label{infTn}
\left({\text{Tr}_R\,\rho_{R}^n}\right)_{\text{eq}} \;{\approx}\;\; \sum_{\sigma \in S_n}\,\dfrac{Z^{R}_{s'_1\beta}\,...\,Z^{R}_{s'_{|\sigma'|}\beta}\,Z^{\bar{R}}_{s_1\beta}\,...\,Z^{\bar{R}}_{s_{|\sigma|}\beta}}{\,(Z^{{R}}_{\beta}Z^{\bar{R}}_{\beta})^n}\,\;+\;\sum_{\pi \in \Pi_n^*}\,{\alpha}_\pi \,|S_\pi| \sum_{\sigma \in S_n/S_\pi } \dfrac{Z^{R}_{q'_1\beta}\,...\,Z^{R}_{q'_{|\sigma_\pi'|}\beta}\,Z^{\bar{R}}_{q_1\beta}\,...\,Z^{\bar{R}}_{q_{|\sigma_\pi|}\beta}}{\,(Z^{{R}}_{\beta}Z^{\bar{R}}_{\beta})^n}\;.
\end{equation}

In the first sum, only the $\sigma = e$ contribution is related to a totally disconnected path integral $(Z^{\bar{R}}_{\beta})^n$, with the dominant bulk saddle being $n$ copies of the Euclidean black hole at inverse temperature $\beta$. The rest of the terms involve a leading contribution of at least a connected geometry coming from $\left(Z_{s\beta_k}\right)$ for $s >1$. The connectivity pattern of these leading geometries is ultimately related to the topology of the replica wormhole that reproduces each term, which also follows the hierarchy between `planar' and `non-planar' contributions \cite{PSSY}. The gravitational replica calculation of the \Rn entropy of the radiation, $(\text{Tr}_R\,\rho_{R}^n)_{\text{grav}}$, seems to be effectively capturing each term of the equilibrium approximation \eqref{eqaprox}. 

For the second sum, the terms are again reminiscent of further suppressed contributions to the gravitational path integral, and they are responsible for restoring the exact unitary description of the equilibrated value. In this case, the extra terms involve the coefficients $\alpha_\pi$ which do not seem to emerge from a boundary path integral like \eqref{PI2}, but rather seem to be combinatorial coefficients prescribed by the long-time average.

\begin{figure}[h]
	\centering
	\includegraphics[width = .73\textwidth]{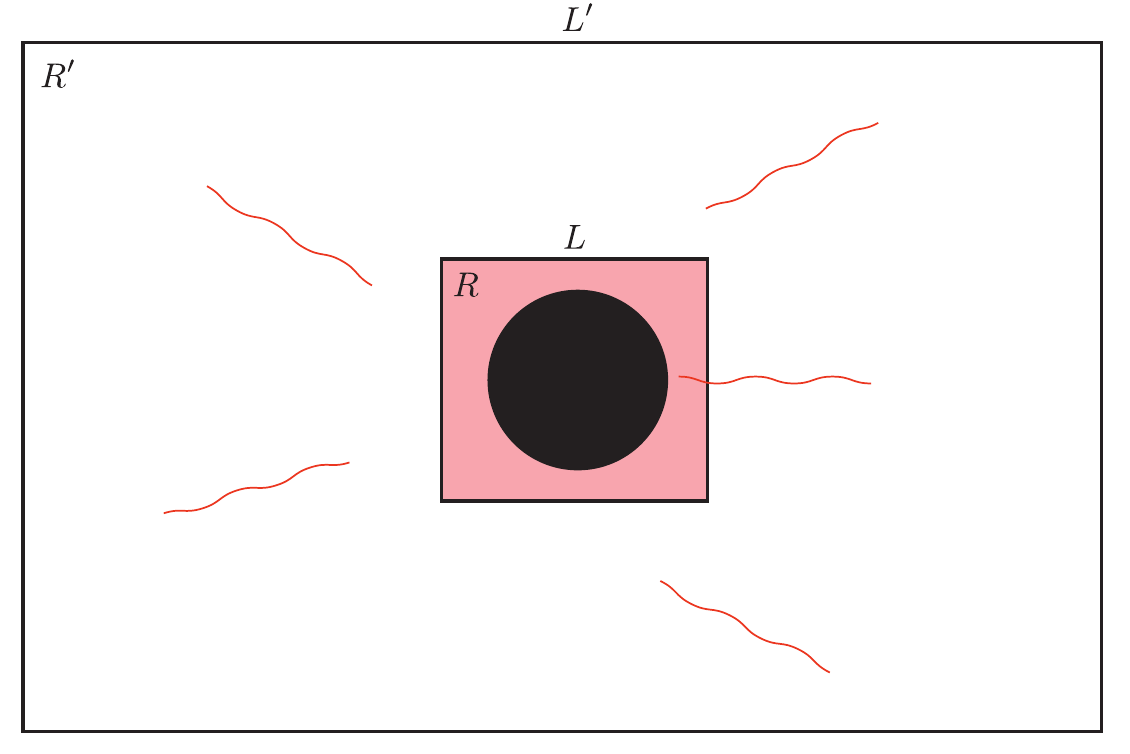}
	\caption{Setup of the model of an evaporating black hole. }
	\label{fig::bhbox}
\end{figure}

We will now consider the situation of an evaporating black hole in this setup. We will first introduce the previous system inside a larger box $R'$ of length $L' \gg L$. Let us couple the small box to the large one by a Hamiltonian $H_{\text{ev}}$ responsible for evaporation, in such a way that radiation escapes slowly compared to the equilibration time $t_{\text{eq}}$ of the small box (see Fig. \ref{fig::bhbox}). This `adiabatic approximation' is a natural assumption for standard black hole evaporation. For example, for a Schwarzschild black hole the evaporation time is of the order of $t_{\text{ev}} \sim \beta \,S_{\text{BH}}$, while the equilibration timescale for the black hole degrees of freedom is of the order of the scrambling time $t_{\text{eq}}\sim \beta \log S_{\text{BH}}$.

Under this approximation, we can divide the full evaporation process in epochs of time $\Delta t$, with $t_{\text{eq}} \ll \Delta t \ll t_{\text{ev}}$. For example, we can consider $\Delta t$ as the time that it takes for the black hole to loose $1\%$ of its initial entropy. At epoch $k$, the system inside the small box will have energy $E_k$, which is a monotonically decreasing function of $k$. We are assuming that the initial energy $E$ is so large that even after many emissions, say when it has lost $99\%$ of its initial entropy, $E_k$ is still large enough to correspond to a stable microcanonical black hole in AdS. 

Consider the radiation subsystem $\text{rad} = R \cup R'$. The \Rn entropies and many observables of the small box rapidly equilibrate, and therefore we can perform a slightly stronger equilibrium approximation
\begin{equation}\label{eqaproxev}
	\text{Tr}_{\text{rad}}\, \rho_{\text{rad},k}^n (t)\; \approx \; \left(\text{Tr}_{\text{rad}}\, \rho_{\text{rad},k}^n\right)_{\text{eq}}\;\approx \;\overline{\text{Tr}_{\text{rad}}\, \rho_{\text{rad},k}^n}\,\Big|_{\text{no ev}}\;\;.
\end{equation}
The long-time average in the last estimation is taken in a different system where no evaporation is allowed. This `eternal' system consists on a black hole inside the box of length $L$, both at inverse temperature $\beta_k$ associated to the energy of the epoch, $E_k$, and no interaction with the larger box $L'$. In this way, we can perform the long-time integration with no risk of loosing track of the black hole subsystem. 

For simplicity, we consider the case in which $H_R = \sum\, \omega \,a_\omega^\dagger \,a_\omega$ is a free Hamiltonian, with the corresponding one-particle Hamiltonian $h = \sum\, \omega \,\ket{\omega}\bra{\omega}$. The one-particle canonical partition functions are defined as $z_n(\beta)\, = \, \sum_\omega e^{-n\beta \omega}$, and $s_n(\beta)$ the corresponding \Rn entropy. Let $\delta s$ be the total number of degrees of freedom emitted in between epochs. Once each emission happens, we can take $L'$ to be arbitrarily large and assume from locality that the emitted quanta do not interact with previously emitted radiation. Under this assumptions, the state $\rho_k$ that approximates the equilibrium properties of $\ket{\Psi(t)}$ at epoch $k$ is the product state
\begin{equation}
	\rho_{k}\, = \, e^{-\beta_1 h\,\delta s}\,\otimes\,...\,\otimes \, e^{-\beta_{k-1} h\,\delta s}\,\otimes \,e^{-\beta_k H_{R}}\,\otimes \, e^{-\beta_k H_{\overline{R}}}\;\;.
\end{equation}
The equilibrium approximation \eqref{eqaproxev} for this density matrix leads to the purity of the radiation
\begin{equation}\label{eqevapsecond}
\left({\text{Tr}_{\text{rad}}\,\rho_{\text{rad},k}^2}\right)_{\text{eq}} \;{\approx}\;\;\;e^{-(S^{R'}_k \,+\, S^R_{\beta_k})}\;+\,e^{-S^{\overline{R}}_{\beta_k}}\,-\,e^{-(S^{R'}_k+S^R_{\beta_k}+S^{\overline{R}}_{\beta_k})}
\end{equation}
where $S^{R'}_k\, = \, \delta s \sum_{j=1}^{k-1}\,s_2(\beta_j)$ is the total second \Rn entropy emitted before the epoch. Similar considerations hold for higher \Rn entropies.

We have shown that the equilibrium approximation for the purity of the radiation \eqref{eqevapsecond} reproduces an exact Page curve. At each epoch, all the terms in \eqref{eqevapsecond} can be formulated as Euclidean path integrals over the corresponding subsystems. In particular, the second term corresponds at leading order to a connected saddle of the gravitational path integral. Remarkably, \eqref{eqevapsecond} yields an exact pure state for the radiation $\rho_R$ at the end of evaporation, even though the adiabatic approximation becomes completely unjustified at the last stages, where the emission timescale is comparable to the scrambling time of the black hole. The last term of \eqref{eqevapsecond} with the corresponding minus sign is responsible for this effect. As we have argued, this minus sign seems to be prescribed from the exact unitary description of the equilibrated purity. Similar considerations hold for higher \Rn entropies.

\section{Conclusions and Outlook}

In this note we have analyzed the equilibration of \Rn entropies under mild assumptions about the chaotic spectral properties of the Hamiltonian of a many-body quantum system. The ergodic long-time average provides a microscopic equilibrium value for the \Rn entropy \eqref{nrenyi} which retains information about the initial state of the system. The averaged quantum noise relative to the long-time value is exponentially suppressed in the microcanonical entropy whenever the subsystem comprises a non-negligible fraction of the full system.

For initial states that excite a large number of energy eigenstates of the microcanonical band, the long-time average of the \Rn entropy is approximated by the microcanonical \eqref{mcav} or the canonical \eqref{mceqnc} equilibrium \textit{ansatze} of \cite{Liu}, with some extra structure that contributes at the level of non-planar permutations. For local systems, each of the terms in the canonical equilibrium approximation for the $n$-th \Rn entropy can be formulated as an Euclidean path integral over $n$ replicas of the system. The extra structure corresponds to path integrals \eqref{PI2} with a more complicated pattern of connectivity between the replicas, which is imposed by the extra projections.

Our results have certain similarities with \Rn entropies for Haar-typical states in the microcanonical band \cite{Lubkin,Lloyd,Page, Page:1993wv, Linden, Nadal}, although we did not assume randomness of the initial state nor of the Hamiltonian. Our considerations apply to atypical initial states which remain atypical for Heisenberg timescales under generic $k$-local time evolution \cite{Verstraete}. It would be interesting to further investigate the scope of our results for states containing a few eigenstates of the Hamiltonian, or even at the level of single chaotic eigenstates, where all the terms in \eqref{nrenyi} become comparable \cite{Dymarsky:2016ntg, Lu:2017tbo, Srednicki3, Fujita:2017pju}.

In the context of AdS/CFT systems describing the semiclassical formation of a black hole that reaches equilibrium with its radiation, the equilibrium approximation yields \eqref{purityfact2} for the purity of the radiation, and \eqref{infTn} for the higher \Rn entropies. These expressions have the same form as the replica calculations in semiclassical gravity, and the connection is strengthen from the point of view of the the CFT path integral that reproduces each term of the equilibrium approximation. We leave the potential identification of the subleading terms of the equilibrium approximation with subleading effects of the semiclassical path integral for future investigation.

The case of the evaporating black hole can also be treated in this formalism under the adiabatic approximation that allows to divide the evaporation process in quasi-equilibrium epochs. A stronger version of the equilibrium approximation yields the purity at each epoch \eqref{eqevapsecond}, where $\beta_k = \beta(t_k)$ is the inverse temperature of the black hole at time $t_k$. Each of the terms in \eqref{eqevapsecond} has also a formulation in terms of an Euclidean path integral over replicas of the whole system, which is given at leading order by a gravitational saddle. The last term in \eqref{eqevapsecond} is responsible for recovering an exact pure state of the radiation at the end of evaporation. It would be interesting to understand whether the semiclassical path integral is able to capture this term with the minus sign in front, and in general the $\alpha_\pi$ coefficients of the corresponding terms for higher \Rn entropies.

\vspace{.5cm}

\noindent{\bf Acknowledgments}  

I am especially grateful to Jos\'{e} Barb\'{o}n for invaluable advice throughout this project and for the detailed feedback on this manuscript, and to Roberto Emparan for many valuable discussions and for the detailed feedback on this manuscript. I would also like to thank Javier Mart\'{i}n, Mikel S\'{a}nchez, Marija Toma{s}evi\'c and Alejandro Vilar for discussions. This work was finished while the author was visiting the Institute of Cosmos Sciences at the University of Barcelona (ICCUB), to whom he is grateful for the warm hospitality. This work is partially supported by the Spanish Research Agency (Agencia Estatal de Investigaci\'{o}n) through the grants IFT Centro de Excelencia Severo Ochoa SEV-2016-0597, and by MINECO through the grant PGC2018-095976-B-C21. The author is supported by the Spanish FPU grant FPU16/00639.

\appendix

\section{Moments of Ergodic Long-Time Averaging}
\label{appendix::A}

In this appendix, we compute the value of several long-time integrals for a Hamiltonian with no rational spectral order.  The integrals correspond to certain complex moments of the homogeneous probability distribution on the ergodic torus $\mathbf{T}^{\mathcal{N}}$. We will define the $n$\textit{-th moment} as the following integral:
\be\label{moment}
	\overline{e^{-it\left(\sum\limits_{s=1}^n\,E_{i_s}\,-E_{j_s}\right)}} = \lim_{T\rightarrow \infty}\,\dfrac{1}{T}\, \int_0^{T}\, \text{d}t\, e^{-it\left(\sum\limits_{s=1}^n\,E_{i_s}\,-E_{j_s}\right)}\; \;,
\ee 
where the value of the $i$ and $j$ indices is unspecified in the range from $1$ to $\mathcal{N}$. 

Generally, the integral as a principal value will only be non-zero when the overall phase vanishes. For instance, the \textit{first moment} yields a delta function
\be
\overline{e^{-it(E_i-E_j)}}\, = \,  \delta^{i}_j\;\;.
\ee

For the \textit{second moment}, the final expression will consists of more terms than the naive permutation of the indices. The reason is that we need to be careful with the over-counting of configuration in which $i_1 = i_2$, since this particular case must yield
\be
\overline{e^{-it(2E_{I}-E_{j_1}-E_{j_2})}}\,   = \,  \delta^{I}_{j_1}\, \delta^{I}_{j_2} \, = \, \delta^{(I}_{j_1}\, \delta^{I)}_{j_2}\;\;, \hspace{1cm}  (i_1 = i_2 = I)\;\;.
\ee

The explicit general formula that captures the above case is
\begin{align}\label{secondmoment}
	\overline{e^{-it(E_{i_1}+E_{i_2}-E_{j_1}-E_{j_2})}}\,   =   \, 2\,\delta^{(i_1}_{j_1}\,\delta^{i_2)}_{j_2}\,\left(1-\,\dfrac{1}{2}\delta^{i_1}_{i_2}\right)\,\;,
\end{align}
where we are not using the convention of summing repeated indices for the last term.

Let us now proceed to the case of the \textit{third moment}. There can be multiple possibilities for coincident $i$-indices. The first one is that only two of them coincide, like for instance $i_1=i_2\neq i_3$. This case will be given by
\begin{align}\label{thirdmequal2}
	\overline{e^{-it(2E_{I}+E_{i_3}-E_{j_1}-E_{j_2}-E_{j_3})}}\,   =   \,\dfrac{3!}{2!}\, \delta^{(I}_{j_1}\,\delta^{I}_{j_2}\,\delta^{i_3)}_{j_3}\;\;, \hspace{1cm}  (i_1 = i_2 = I\neq i_3)\;\;,
\end{align}
where the prefactor arises from the double-counting of the permutations that swap the equal $I$-indices. It can also be the case that all three $i$-indices coincide. In that case, the prefactor must be different
\begin{align}\label{thirdmequal3}
	\overline{e^{-it(3E_{I}-E_{j_1}-E_{j_2}-E_{j_3})}}\,   =   \, \delta^{I}_{j_1}\,\delta^{I}_{j_2}\,\delta^{I}_{j_3}\,= \,\delta^{(I}_{j_1}\,\delta^{I}_{j_2}\,\delta^{I)}_{j_3} \;\;, \hspace{1cm}  (i_1 = i_2 = i_3 = I)
\end{align}

One can easily check that the fomula that encapsulates all the above cases is given by
\be\label{thirdmoment}
\overline{e^{-it(E_{i_1}+E_{i_2}+E_{i_3}-E_{j_1}-E_{j_2}-E_{j_3})}}\,   =     \, 3!\,\delta^{(i_1}_{j_1}\delta^{i_2}_{j_2}\delta^{i_3)}_{j_3}\,\left(1\,-\,\dfrac{1}{2}\delta_{i_2}^{i_1}\,-\,\dfrac{1}{2}\delta_{i_3}^{i_2}\,-\,\dfrac{1}{2}\delta_{i_1}^{i_3}\,+\,\dfrac{2}{3}\,\delta_{i_2}^{i_1}\delta_{i_3}^{i_2}\right) \;.
\ee

The generalization for general $n$ is straightforward once we do a little bit of combinatorics. The integral will have the form 
\begin{align}\label{nmoment}
	\overline{e^{-i\left(\sum\limits_{s=1}^n\,E_{i_s}\,-E_{j_s}\right)}}\,  =   \, n!\,\delta^{(i_1}_{j_1}...\delta^{i_n)}_{j_n}\,\sum_{\pi \in \Pi_n}\,\alpha_{\pi}\,\prod_{B\in \pi}\prod_{a,b \in B}\delta^{i_{a}}_{i_{b}}\;.
\end{align}

Let us explain this formula in more detail. The prefactor just represents the naive ways of  assigning the $i$ indices to the $j$ indices. The sum is over the set of partitions of the set $\mathbf{N}_n = \lbrace 1,2,...,n\rbrace $, denoted by $\Pi_n$. A given partition $\pi = \lbrace{B_1,...,B_k\rbrace}$ just represents that indices in each of the blocks $B_i$ are equal. For example, for $n=7$, the partition $\pi = \lbrace{\lbrace 3,4\rbrace, \lbrace 2,5\rbrace, \lbrace 1,6, 7\rbrace\rbrace}$ represents that $i_3 = i_4$, $i_2 = i_5$ and $i_1 = i_6 = i_7$. This is the reason for the two products, the first one being over the blocks of the partition, and the second one over indices in each of the blocks, which must be set equal with the delta function. The number of terms in the sum is therefore the number of partitions of a set of $n$ elements, which is called the Bell number $\mathcal{B}_n$. This number grows super-exponentially for large $n$.

Now, we need to fix the $\alpha_\pi$ coefficients in front of each of the terms in the sum. We will do this recursively. First of all, the trivial partition $\pi_e = \lbrace{\lbrace{1}\rbrace, \lbrace{2}\rbrace,...,\lbrace{n}\rbrace\rbrace}$ that represents the counting when all indices are different will have $\alpha_{\pi_e} = 1$. For a generic partition $\pi = \lbrace{B_1,...,B_k\rbrace}$, the total number of configurations that should be counted is just given by the ways to arrange $n$ elements on different boxes of sizes $|B_i| = n_i$, that is, $n!/\left(n_1!\,... n_k!\right)$. The reason is that all permutations that can be related by a permutation of the identical indices leave the partition invariant. 

The coefficient $\alpha_{\pi}$ will only depend on the size of the boxes $n_1, ..., n_k$. That is, it depends on the equivalence class of the partition, where two partitions are equivalent if their $B$'s have the same number of elements. These equivalence classes are just given by the partitions of the natural number $n$, which are all the ways to decompose $n$ as a sum of positive integers. For instance, the trivial partition $\pi_e$ is the only element of the class $n = 1+1+...+1 \equiv [1^n]$. All the partitions which have a pair, like $\lbrace{\lbrace{1,2}\rbrace, \lbrace{3}\rbrace,...,\lbrace{n}\rbrace\rbrace}$ or $\lbrace{\lbrace{7,56}\rbrace ,\lbrace{1}\rbrace, \lbrace{2}\rbrace, ...,\lbrace{n}\rbrace\rbrace}$, will belong to the class $n = 2+1+...+1 \equiv [2,1^{n-2}]$. Partitions that consist of a triplet will belong to the class $n = 3+1+...+1 \equiv [3,1^{n-3}]$. Partitions with two distinct pairs will be of the class $n = 2+2+1+...+1 \equiv [2^2,1^{n-4}]$, and so on.

Let us start from all partitions of the class $[2,1^{n-2}]$, that is, $\pi$ consists of just one pair. Already the trivial partition $\pi_e$ accounts for $n!$ cases of this kind. All other terms in \eqref{nmoment} do not contribute since we are assuming that only two indices are equal, and therefore only the actual partition and possibly finer partitions can only contribute. We therefore have that
\be
\dfrac{n!}{2!} = n! \left(1\,+\,\alpha_{[2,1^{n-2}]}\right) \,\Rightarrow\,\alpha_{[2,1^{n-2}]} = - \dfrac{1}{2} \;.
\ee

For triplets, i.e. partitions in the class $[3,1^{n-3}]$,  we have that the trivial partition term counts $n!$ such cases, but some of the $[2,1^{n-2}]$ partitions can also contribute. In particular, there are ${3\choose 2}$ such partitions that contribute. We therefore have 
\be
\dfrac{n!}{3!}  = n! \left(1\,+\,{3\choose 2}\alpha_{[2,1^{n-2}]}\,+\,\alpha_{[3,1^{n-3}]}\right) \Rightarrow\,\alpha_{[3,1^{n-3}]} = \dfrac{2}{3}\;.
\ee

For two doublets, i.e. partitions in the class $[2^2,1^{n-4}]$, we would have
\be
\dfrac{n!}{2!2!}  = n! \left(1\,+\,2\alpha_{[2,1^{n-2}]}\,+\,\alpha_{[2^2,1^{n-4}]}\right) \Rightarrow\,\alpha_{[2^2,1^{n-4}]} = \dfrac{1}{4}\;.
\ee

For quadruplets, i.e. partitions in the class $[4,1^{n-4}]$, the result would be
\be
\dfrac{n!}{4!}  = n! \left(1\,+\,{4\choose 2}\alpha_{[2,1^{n-2}]}\,+\,{4\choose 3}\alpha_{[3,1^{n-3}]}\,+\,{4\choose 2}\,\alpha_{[2^2,1^{n-4}]}\,+\,\alpha_{[4,1^{n-4}]}\right) \Rightarrow\,\alpha_{[4,1^{n-4}]} = -\dfrac{17}{8}\;.
\ee

For a general partition $\pi = \lbrace{B_1,...,B_k\rbrace}$, only when we know all the coefficients for finer partitions $\pi' < \pi$ will we be able to solve for the coefficient through the following condition
\be
\dfrac{n!}{n_1!\,...\,n_k!}  = n! \sum_{\pi ' \leq \pi }\,\alpha_{\pi'}\,  \hspace{1cm}\Rightarrow \hspace{1cm}\,\alpha_{\pi} \, = \,  \dfrac{1}{n_1!\,...\,n_k!}\,-\,\sum_{\pi ' < \pi }\,\alpha_{\pi'}\,\;.
\ee

\begin{table}[H]
	\centering
	\begin{tabular}{|  c  |  c  |  c  |  c  |  c  |  c  |  c  |  c  |}
		\hline
		$\pi$ & $[1^n]$ & $[2,1^{n-2}]$ & $[3,1^{n-3}]$ & $[2^2,1^{n-4}]$ & $[4,1^{n-4}]$ & $[3,2,1^{n-5}]$ & $[5,1^{n-5}]$ \\
		\hline
		\rule{0pt}{20pt}$\alpha_\pi$ & $1$ &$-\dfrac{1}{2}$ & $\dfrac{2}{3} $ & $\dfrac{1}{4} $ & $-\dfrac{17}{8} $ & $\dfrac{5}{12} $ & $\dfrac{7}{15} $  \\[10pt]
		\hline
	\end{tabular}
	\caption{First few values of $\alpha_\pi$ for the finest partitions.}
	\label{tablepis}
\end{table}

For example, let us recover \eqref{thirdmoment} from this general formula, by substituting $n=3$. The Bell number is in this case $\mathcal{B}_3 = 5$. The list of partitions of $\mathbf{N}_3$ is
\begin{gather}
	\pi_e = \lbrace{\lbrace{1}\rbrace, \lbrace{2}\rbrace,\lbrace{3}\rbrace\rbrace}\, \in [1^3]\;\;, \\
	\lbrace{\lbrace{1,2}\rbrace, \lbrace{3}\rbrace\rbrace}\, \in [2,1]\;\;, \\
	\lbrace{\lbrace{2,3}\rbrace, \lbrace{1}\rbrace\rbrace}\, \in [2,1]\;\;, \\
	\lbrace{\lbrace{1,3}\rbrace, \lbrace{2}\rbrace\rbrace}\, \in [2,1]\;\;, \\
	 \lbrace{\lbrace{1,2,3}\rbrace\rbrace}\, \in [3] \label{partitions3}\;\;.
\end{gather}

The coefficient for all these partitions can be read from Table \ref{tablepis}. The formula \eqref{nmoment} then particularizes exactly to \eqref{thirdmoment}.

The case for $n=4$ will consist of $\mathcal{B}_4 = 15$ terms, which can be read from the table
\begin{gather}
	\overline{e^{-i\left(\sum\limits_{s=1}^4\,E_{i_s}\,-E_{j_s}\right)}}=   \, 4!\,\delta^{(i_1}_{j_1}\delta^{i_2}_{j_2}\delta^{i_3}_{j_3}\,\delta^{i_4)}_{j_4}\,\left(1\,-\,\dfrac{1}{2}\delta_{i_2}^{i_1}\,-\,\dfrac{1}{2}\delta_{i_3}^{i_1}\,-\,\dfrac{1}{2}\delta_{i_4}^{i_1}\,-\dfrac{1}{2}\delta_{i_2}^{i_3}\,-\dfrac{1}{2}\delta_{i_2}^{i_4}\,-\dfrac{1}{2}\delta_{i_3}^{i_4}\,\right. \nonumber \\ \left. +\,\dfrac{2}{3}\,\delta_{i_1}^{i_2}\delta_{i_2}^{i_3}\,+\,\dfrac{2}{3}\,\delta_{i_2}^{i_1}\delta_{i_4}^{i_2}\,+\,\dfrac{2}{3}\,\delta_{i_3}^{i_1}\delta_{i_4}^{i_3}+\,\dfrac{2}{3}\,\delta_{i_3}^{i_2}\delta_{i_4}^{i_3}\,+\,\dfrac{1}{4}\,\delta^{i_1}_{i_2}\delta^{i_3}_{i_4}\,+\,\dfrac{1}{4}\,\delta^{i_1}_{i_3}\delta^{i_2}_{i_4}\,+\,\dfrac{1}{4}\,\delta^{i_1}_{i_4}\delta^{i_2}_{i_3}\,-\,\dfrac{17}{8}\,\delta^{i_1}_{i_2}\delta^{i_2}_{i_3}\delta^{i_3}_{i_4}\right) \;\;.
\end{gather}

The case $n=5$ already has $\mathcal{B}_5 = 52$ terms, so writing \eqref{nmoment} in an explicit form becomes really tedious.

\bibliographystyle{style}
\bibliography{eqap.bib}
\end{document}